\documentclass[preprint,showpacs,preprintnumbers,amsmath,amssymb]{revtex4-1}

\usepackage{graphicx}% Include figure files
\usepackage{dcolumn}% Align table columns on decimal point
\usepackage{bm}% bold math

\begin{document}

\preprint{}

\title{Bose-like condensation of Lagrangian particles and higher-order statistics
in passive scalar turbulent advection}% Force line breaks with \\

\author{Thierry Dombre}
\affiliation{
Laboratoire de Spectrom\'{e}trie Physique, CNRS-Universit\'{e} Joseph Fourier, BP87,
38402 Saint-Martin d'H\`{e}res C\'{e}dex, France
}

\date{\today}

\begin{abstract}
We establish an hitherto hidden connection between zero modes and instantons in the
context of the Kraichnan model for passive scalar turbulent advection, that relies on
the hypothesis that the production of strong gradients of the scalar is associated
with Bose-like condensation of Lagrangian particles. It opens the way to the computation
of scaling exponents of the $N$-th order structure functions of the scalar by techniques
borrowed from many-body theory.  To lowest order of approximation, scaling exponents are
found to increase asymptotically as $\log N$ in two dimensions.
\end{abstract}

\pacs{47.27.eb, 02.50.-r}

\maketitle

There has been in the last two decades great progress in the quantitative
understanding of anomalous scaling in fluid turbulence. In the framework
of the Kraichnan model for passive scalar advection, the emergence of
anomalous scaling has been traced
to the existence of statistical integrals of motion showing up in the
evolution of Lagrangian fluid particles (see \cite{FGV01} for a review).
For any finite number of particles, the conserved quantities are functions of
the interparticle separations that are statistically preserved as the particles
are transported by the random flow and scale as a power law with respect to the
mean radius of the cloud of particles. The scaling exponent of these so-called
zero modes (for reasons to get obvious below) depend in a nonlinear way on
the number $N$ of particles and it turns out that the scaling behaviour of the
$N$-th order structure functions of the advected scalar
$T_{N}(r)=\langle (\theta(x+r)-\theta(x))^{N}\rangle$
(for $N$ even) is dominated by the irreducible zero mode of corresponding order
and lowest positive scaling dimension. Although some specific features
of the Kraichnan model are needed to establish on a firm mathematical ground the
existence of zero modes, this mechanism is believed to be robust and relevant for
transport by generic turbulent flows.

On the other hand, for large values of $N$, one generally expects $N$-point
structure functions to be controlled by rare events which can be captured
by the instanton formalism first introduced in the field of particle physics
(see, e.g., Coleman \cite{Coleman77}) and adapted later on for turbulence
by Falkovich, Migdal and co-workers \cite{FKLM96}, at about the same time as
the zero modes breakthrough took place.
In this second approach, one looks for configurations of the advecting velocity field
(the random noise in the Kraichnan model) of optimal statistical weight leading to a
prescribed value of the scalar increment at short scales. When the dynamics in the inertial
range is scale invariant, one may focus on self-similar instantons as shown
in the framework of shell models of turbulence \cite{BDDL99,DDG00}. Those objects
yield the best picture of the singular scaling fluctuations
$\delta_{r}\theta (x) = \theta (x+r)-\theta (x) \sim r^{h(x)}$, which are at the basis
of the phenomenological multifractal description of turbulence introduced by Parisi
and Frisch sometime ago \cite {PF85}. In the multifractal model, the probability of a
$h-$fluctuation occurring is given by $P_{r}(h)\sim r^{s(h)}$, where the function $s(h)$
can be interpreted as the extinction rate of the singular fluctuation as it cascades towards
small scales.
Structure functions of any order may be computed by averaging over
the probability $P_{r}(h)$ the appropriate power of the fluctuating field.
It is then easily recognized that $T_{N}(r) \propto \int dh \, r^{(s(h)+Nh)}$ is a
scaling function of $r$ in the inertial range, of the form  $T_{N}(r) \sim r^{\zeta_{N}}$
with $\zeta_{N}=\min_{h}[s(h)+Nh]$. Scaling exponents $\zeta_{N}$ are therefore linked
by a Legendre transform to the function $s(h)$, whose instanton formalism is able
to produce a first estimate (dressing of the instanton by fluctuations has to be considered
in order to improve the result).

So far, the connection between these two descriptions of the origin of intermittency in
stochastic turbulent systems has remained elusive. Computations of zero modes and their scaling
exponents in the framework of the Kraichnan model were mostly done using perturbative methods
around limiting values of its parameters for which anomalous scaling disappears (high space
dimensionality, small or large roughness degree of the advecting velocity field). Those methods
fail to capture the scaling exponents in the large $N$ nonperturbative domain (reached, roughly
speaking, when $N$ gets larger than the space dimensionality $d$). Lagrangian numerical methods as
introduced in \cite{FMV98} offer an alternative strategy for computing zero modes. But in practice,
due to the rapid increase of the time of integration of the equations of motion
needed to reach the required accuracy as the number $N$ of Lagrangian particles increases, they
are bound to explore moderate values of statistical orders.

Alternatively, there has been some attempts at developing the instanton formalism
for the Kraichnan model \cite{Chertkov97, BL98}. The most convincing one was carried out by Balkovsky
and Lebedev \cite{BL98}. In order to get rid of soft modes (a common source of trouble in path
integral formulations and their treatment by saddle-point approximations), physically associated with
slow modulations of the strength and the orientation of the local strain matrix, these authors wrote
down an effective theory concentrating on the evolution of interparticle distances, which could be
then solved only by considering that the space dimensionality $d$ is large. The sequence of bold
simplifying assumptions introduced in \cite{BL98} leaves little hope for understanding along this
way the link between zero modes and instantons in the large $N$ limit.

It is the main purpose of this Letter to bridge the gap between these two cornerstones of the modern
explanation of the emergence of anomalous scaling in turbulent advection. In order to do so, we shall
rely on the crucial observation that the production of large scalar differences at microscopic
scales requires correlated motion of Lagrangian particles coming from different regions of space.
In a quantum analogy, this will be achieved provided particles, seen as bosons, condense in the same state.
Standard methods developed in condensed matter physics for Bose condensation will allow us to reduce in the
large $N$ limit the $N$-body problem one is facing in the zero mode approach to an effective one-particle
problem involving both diffusion and advection in the mean velocity field created by the remaining $N-1$
particles. Estimates for scaling exponents of zero modes can then be inferred from the ground state energy
of the Bose condensate. Furthermore, it turns out that the equations fixing its wave-function
can be mapped exactly onto the ones coming out from an instanton analysis applied to the original stochastic
equations governing the evolution of the full scalar field $\theta(x,t)$ in $d$-dimensional space.
A numerical study of these equations, performed in the present work at $d=2$, leads to the prediction of
scaling exponents $\zeta_{N}$ growing like $\log N$ at large $N$. This is at odds with the common belief
\cite{Yakhot97, Chertkov97} that the $\zeta_{N}$'s should saturate at large $N$, due to the statistical
preeminence of fronts carrying
finite discontinuities of the scalar like in Burgers equation. We do not know whether the logarithmic growth
of scaling exponents found in our approach is a robust feature which will survive upon the inclusion of
fluctuations around the Hartree-Fock description of the Bose condensate presented in this Letter. We note
however the absence of a definitive theoretical argument in favor of saturation of the scaling exponents
and leave the resolution of this discrepancy open for further work.\\

In the inertial range (where forcing and dissipation can be neglected), the equation of motion of
the Kraichnan model
\begin{equation}\label{advection}
    \partial_{t}\theta({\bf r},t) + {\bf v}({\bf r},t).{\bf \nabla}\theta({\bf r},t)=0,
\end{equation}
just describes passive advection of the scalar quantity $\theta({\bm r},t)$ in a random Gaussian
incompressible velocity field ${\bf v}({\bf r},t)$ whose two-point correlations behave like
\begin{equation}\label{correl}
\langle v_{i}({\bf r},t) v_{j}({\bf r'},t')\rangle = 2 \delta (t-t')(D_{0}\delta_{ij}-d_{ij}({\bf r}-{\bf r'})),
\nonumber
\end{equation}
with $d_{ij}({\bf r})=r^{\xi}\left((d-1+\xi)\delta_{ij}-\xi \frac{r_{i}r_{j}}{r^{2}}\right)$.
The exponent $\xi $, which fixes the way velocity differences scale at short distances and ranges between
0 and 2, together with the space dimensionality $d$, are the two physically important parameters of the
Kraichnan model. In eq.~(\ref{correl}), spatial indices $i$ and $j$ run between $1$ and $d$, and the incompressibility
of the velocity field is warranted by the following property of the matrix $d_{ij}({\bf r})$ :
$\nabla_{i}d_{ij}({\bf r})=0, \forall j $.

Let us now consider $N$ fluid particles, close to each other at initial time $t=0$, and transported at
later times in various realizations of the Kraichnan velocity field. We define
${\cal P}_{N}({\bf r}_{1},{\bf r}_{2},\ldots {\bf r}_{N},t)$ as the PDF of their positions
${\bf r}_{1},{\bf r}_{2},\ldots {\bf r}_{N}$ at time $t$ (to keep notations simple, we skip the
reference to the initial positions of Lagrangian particles in the arguments of ${\cal P}_{N}$.
We shall also use in the following $\underline{\bf r}$ as a short-hand notation for
$({\bf r}_{1},{\bf r}_{2},\cdots,{\bf r}_{N})$).
In the translation-invariant sector, ${\cal P}_{N}$ evolves as \cite{FGV01}
\begin{equation}\label{dynamics}
   \frac{d{\cal P}_{N}}{dt}= \left(-\sum_{n<m}d_{ij}({\bf r}_{nm})\nabla_{r^{i}_{n}}\nabla_{r^{j}_{m}}\right){\cal P}_{N}
   \equiv {\cal M}_{N}{\cal P}_{N},
\end{equation}
where particles are labelled by integers $n$ or $m$ and vectors
${\bf r}_{nm}={\bf r}_{m}-{\bf r}_{n}$ denote their relative positions.

The operator ${\cal M}_{N}$ is of dimension $\xi -2$ with respect to length rescaling. As a
consequence, there exist non trivial solutions to the equation ${\cal M}_{N} f =0$
(or zero modes), that are scaling functions of positive dimension $\zeta$ (\textit{i.e.}, such that
$f(\lambda \underline{\bf r})= \lambda^{\zeta} f(\underline{\bf r})$). Since ${\cal M}_{N}$
is a self-adjoint operator, one easily deduces that the statistical average of zero modes
(defined in the translation-invariant sector as $\langle f \rangle(t)= \int f(\underline{\bf r})
{\cal P}_{N}(\underline{\bf r},t) d^{d}r_{1}\cdots d^{d}r_{N-1}$
is conserved by the dynamics. It was shown (see again the review paper \cite{FGV01} for more
details) that zero modes are formally present
in the $N$-point equal-time correlation function of the scalar field and that the exponent $\zeta_{N}$
can be identified with the smallest possible value of scaling dimensions of irreducible
zero modes of ${\cal M}_{N}$, \textit{i.e.}, those modes which do not belong to the kernel of
sub-operators ${\cal M}_{p}$ with $p <N$.

In order to make the computation of $\zeta_{N}$ more amenable to techniques of many body theory,
we first relax the constraint of translation invariance by adding a supplementary particle at
position ${\bf r}_{N+1}$ and write ${\cal P}_{N}$ as
${\cal P}_{N}(\underline{\bf r},t)
= \int \widetilde{\cal P}_{N}({\bf r}_{1}-{\bf r}_{N+1},\cdots, {\bf r}_{N}-{\bf r}_{N+1},t)
\;d^{d}r_{N+1}$. The new function $\widetilde{\cal P}_{N}$, as the PDF
of relative positions in a cloud of $N+1$ particles, obeys the equation
$\frac{d\widetilde{\cal P}_{N}}{dt}= \widetilde{\cal M}_{N}\widetilde{\cal P}_{N}$,
where
%\begin{eqnarray}
%\widetilde{\cal M}_{N}&=& \frac{1}{2}\sum_{n,m=1}^{N}
%\left\{d_{ij}({\bf r}_{N+1n})+d_{ij}({\bf r}_{N+1m})-d_{ij}({\bf r}_{nm})\right\}
%\nonumber \\
%&& \times \nabla_{r^{i}_{n}} \nabla_{r^{j}_{m}}
%\end{eqnarray}
\begin{equation}
\widetilde{\cal M}_{N}= \frac{1}{2}\sum_{n,m=1}^{N}
\left\{d_{ij}({\bf r}_{N+1n})+d_{ij}({\bf r}_{N+1m})-d_{ij}({\bf r}_{nm})\right\}
 \times \nabla_{r^{i}_{n}} \nabla_{r^{j}_{m}}
\end{equation}
is nothing but ${\cal M}_{N+1}$ expressed in the referential frame of the $(N+1)$-th particle
(from now on, we shall put without loss of generality ${\bf r}_{N+1}={\bf 0}$ and note
${\bf r}_{N+1n}= {\bf r}_{n}-{\bf r}_{N+1}$ as ${\bf r}_{n}$).
It follows that zero modes of ${\cal M}_{N}$ are also zero modes of
$\widetilde{\cal M}_{N}$. Another advantage of this new formulation of the problem lies in the
fact that interactions or correlations between the particles in the original cloud
and the central one will be taken into account exactly, even in the approximation schemes
we shall be obliged to introduce later on to make progress. This is close in spirit to the
Bethe-Peierls method in statistical mechanics, known to give a better description of
local ordering in condensed phases than basic mean field theory.

We then take advantage of the scaling properties of the operator $\widetilde{\cal M}_{N}$
to switch towards a representation incorporating in a natural way the average expansion of
length scales implied by the dynamics (\ref{dynamics}). We rewrite $\widetilde{\cal P}_{N}$
as $\widetilde{\cal P}_{N}(\underline{\bf r},t) =t^{-\frac{Nd}{\gamma}}
\Phi_{N}(\underline{\bf r}\,t^{-\frac{1}{\gamma}},\ln t)$ with $\gamma = 2-\xi$.
After defining new space and time variables ${\bm \rho}_{n}={\bf r}_{n} t^{-\frac{1}{\gamma}}$
and $\tau = \ln t$, the evolution of the PDF takes the final form~:
\begin{equation}\label{def LN}
    \frac{d{\Phi}_{N}}{d\tau}= \left\{\widetilde{\cal M}_{N} +\frac{\Lambda_{N}}{\gamma}
   +\frac{Nd}{\gamma}\right\}{\Phi}_{N}\equiv {\cal L}_{N}{\Phi}_{N},\nonumber
\end{equation}
where $\Lambda_{N}=\sum_{n}\rho_{n}^{i}\nabla_{\rho_{n}^{i}}$ is the generator of scale
transforms in the $Nd$-dimensional space of configurations and $\widetilde{\cal M}_{N}$
 is now expressed in terms of the rescaled position variables
${\bm \rho}_{n}$. Unlike $\widetilde{\cal M}_{N}$, the operator ${\cal L}_{N}$ is not
self-adjoint. Since $^{t}{\cal L}_{N}=\widetilde{\cal M}_{N} -\frac{\Lambda_{N}}{\gamma}$,
we deduce that zero modes are left eigenvectors of ${\cal L}_{N}$ (of eigenvalue
$-\frac{\zeta}{\gamma}$ if $\zeta$ denotes their scaling dimension). A more detailed
analysis reveals that zero modes lie at the top of a ``tower'' of left eigenvectors
of ${\cal L}_{N}$ of eigenvalue $-\frac{\zeta}{\gamma}-k$ where $k$ is an integer (those
eigenvectors are obtained as linear combinations of the slow modes discovered in \cite{BGK98}).
Corresponding right eigenvectors of ${\cal L}_{N}$ are of finite norm.
It should be noted that zero modes are also right eigenvectors of ${\cal L}_{N}$, and as such
give rise to a positive part in the spectrum of this operator consisting of eigenvalues of the form
$\frac{\zeta}{\gamma}+\frac{Nd}{\gamma}+ k$ (with $k$ an integer). However this part of the spectrum
cannot take part in the time evolution of the PDF since it is built from non normalizable right
eigenvectors, and will be of no concern in the following discussion.

We are now in a good position for catching an estimate for $\zeta_{N}$ by variational approach.
We can see indeed $\zeta_{N}$ as the minimum of the functional $\langle\Psi|-\gamma{\cal L}_{N}|\Phi\rangle$
for any pair of left and right irreducible states $\langle\Psi|$ and $|\Phi\rangle$ of unit overlap
$\langle\Psi|\Phi\rangle$.
Irreducibility of the right state $|\Phi\rangle$ (which is enough to project the whole variational
procedure on the desired Hilbert space) is easily enforced by the condition
$\int \Phi(\underline{\bm \rho}) \;d^{d}\rho_{n}=0, \forall n \leq N$. The latter is indeed just a way of
stating that the trial state $|\Phi\rangle$ does not belong to the dual space of the kernels of
sub-operators ${\cal M}_{p}$ with $p <N$. We assume Bose condensation
of particles in a dumbbell-like geometry, and restrict our attention to variational states
of the form
\begin{equation}\label{orbitals}
    \Phi(\underline{\bm \rho})=\prod_{n=1}^{N}\varphi_{0}({\bm \rho_{n}}), \;\;
    \Psi(\underline{\bm \rho})=\prod_{n=1}^{N}\psi_{0}({\bm \rho_{n}}),
\end{equation}
where both orbitals $\varphi_{0}$ and $\psi_{0}$ are odd with respect to inversion of coordinates
along the dumbbell axis and invariant with respect to rotations around this axis.
Minimizing $\langle\Psi|-\gamma{\cal L}_{N}|\Phi\rangle$ under the constraint
$\langle\psi_{0}|\varphi_{0}\rangle=1$ leads
to the following conditions for $\varphi_{0}$ and $\psi_{0}$~:
\begin{equation}\label{equation-orbitales}
    -l_{N}\,\varphi_{0} = \mu_{N}\,\varphi_{0},\;\; -^{t}l_{N}\,\psi_{0}= \mu_{N}\,\psi_{0},
\end{equation}
where $\mu_{N}$ is a Lagrangian multiplier and the one-particle operator $l_{N}$ reads
\begin{equation}\label{hamiltonien-effectif}
    l_{N}=d_{ij}({\bm \rho})\nabla_{i}\nabla_{\!j}
    +[V_{i}({\bm \rho})+\frac{\rho_{i}}{\gamma}]\nabla_{i}+\frac{d}{\gamma},
\end{equation}
with
\begin{equation}\label{Vitesse}
    V_{i}({\bm \rho}) = (N-1)\! \int \! d_{ij}({\bm \rho}-{\bm \rho'})\nabla_{\!j}\psi_{0}({\bm \rho'})
    \,\varphi_{0}({\bm \rho'})\,d^{d}\!\rho'.
\end{equation}
Particles are seen to condense in the state of lowest ``energy'' $\mu_{N}$ of an effective Hamiltonian
involving both diffusion and advection in an incompressible velocity field ${\bf V}(\bm \rho)$ defined
by eq.~(\ref{Vitesse}), which expresses on an averaged way the interactions with other particles and
adds to the radial velocity component issuing from the continuous rescaling of lengths. Note that the
presence of a diffusion term in the effective Hamiltonian $l_{N}$ is essential to ensure the existence
of non trivial and physically meaningful solutions to eq.~(\ref{equation-orbitales}). We obtain the
following upper bound for $\zeta_{N}$~:
\begin{equation}\label{exposant}
  \zeta_{N}^{(0)}=\gamma \left\{N\mu_{N} - \frac{N(N-1)}{2}
  \int \psi_{0}({\bm \rho}) \psi_{0}({\bm \rho'})\,d_{ij}({\bm \rho}-{\bm \rho'})
 \nabla_i \varphi_{0}({\bm \rho})\nabla_{\!j} \varphi_{0}({\bm \rho'})\,
  d^{d}\!\rho \, d^{d}\!\rho'\right\}.
\end{equation}
%\begin{eqnarray}\label{exposant}
%  \zeta_{N}^{(0)}&=&\gamma \{N\mu_{N} - \frac{N(N-1)}{2}
%  \int \psi_{0}({\bm \rho}) \psi_{0}({\bm \rho'})\,d_{ij}({\bm \rho}-{\bm \rho'})\nonumber\\
% &&\nabla_i \varphi_{0}({\bm \rho})\nabla_{\!j} \varphi_{0}({\bm \rho'})\,
%   d^{d}\!\rho \, d^{d}\!\rho'\}.
%\end{eqnarray}
We get another interesting relation from our theory by treating $N$ as a continuous variable.
Differentiation of $\mu_{N}\!\!=\!\!-\langle\psi_{0}|l_{N}|\varphi_{0}\rangle$
with respect to $N$ leads to the result $\frac{d\mu_{N}}{dN}\!\!=\!\!-\frac{1}{N-1}\frac{d{\cal S}_{N}}{dN}$
where we define ${\cal S}_{N}\!\!=\!\!-\frac{(N-1)^{2}}{2}\int \psi_{0}({\bm \rho}) \psi_{0}({\bm \rho'})\,
d_{ij}({\bm \rho}-{\bm \rho'})\nabla_i \varphi_{0}({\bm \rho})
\nabla_{\!j} \varphi_{0}({\bm \rho'})\,d^{d}\!\rho \, d^{d}\!\rho'$.
Considering $S_{N}$ as a function of $\mu$ rather than $N$ (via $\mu_{N}$), we can rewrite in the large $N$ limit
the expression obtained before for $\zeta_{N}^{(0)}$ as
\begin{equation}\label{exposant-simple}
  \zeta_{N}^{(0)}=\gamma \{N\mu_{N}+{\cal S}(\mu_{N})\}\;\mbox{with}\;{\cal S}'(\mu_{N})=-N.
\end{equation}
In other words, up to the pre-factor $\gamma$ converting time scales into spatial ones, the scaling exponent
$\zeta_{N}$ is nothing but the Legendre transform of the function ${\cal S}(\mu)$. We have therefore recovered,
at this level of approximation, the phenomenological content of the multifractal model of Parisi and Frisch.\\

Let us now show how the preceding results also arise from the instanton approach. In order to do so,
we shall follow the formulation of instanton theory introduced in \cite {BDDL99, DDG00}, though in the
restricted framework of shell models. We go back to eq.~(\ref{advection}) and first eliminate global
sweeping effects by adopting a quasi-Lagrangian description. This amounts to defining all the fields
in a frame whose origin moves with the fluid and transforms eq.~(\ref{advection}) into
\begin{equation}\label{QL-advection}
    \partial_{t}\theta({\bf r},t) + [{\bf v}({\bf r},t)-{\bf v}({\bf 0},t)].{\bf \nabla}\theta({\bf r},t)=0,
\end{equation}
upon appropriate redefinition of coordinates and fields. We then go from the Stratonovich prescription
underlying the above stochastic equation to the It\^{o} convention. According to standard rules of
stochastic calculus \cite {Gardiner85}, eq.~(\ref{QL-advection}) becomes
\begin{equation}\label{Ito-advection}
    \partial_{t}\theta + [{\bf v}({\bf r},t)-{\bf v}({\bf 0},t)].{\bf \nabla}\theta
    - d_{ij}({\bf r})\nabla_i \nabla_j \theta =0.
\end{equation}
The It\^{o} form of the equation has the merit of making explicit the mixing of the scalar due to short-range
fluctuations of the velocity-field, in the form of an effective eddy diffusivity varying with the relative
length scale in the local frame. Due to this eddy diffusivity, the formation of fronts in the scalar field
cannot be anymore arbitrarily slowed down as formally allowed by eq.~(\ref{QL-advection}). Furthermore, the
evolution of the scalar field in this representation becomes a true Markov process, which facilitates the
writing of its PDF as a path integral over various realizations of the noise. Indeed, following for instance
\cite{FKLM96}, it is rather easy to establish that the probability of reaching a prescribed scalar
configuration $\theta_{f}({\bf r})$ at time $t_{f}$, being given an initial one $\theta_{in}({\bf r})$ at time
$t_{in}$, can be expressed in terms of the Martin-Siggia-Rose path integral
$\int {\cal D}\theta \,{\cal D}\!p \;\exp -S$, with the action
\begin{equation}\label{MSR-action}
  S=\int_{t_{in}}^{t_{f}} dt \left\{\int \! d^{d}r\, ip({\bf r},t).
  [\partial_{t}\theta - d_{ij}({\bf r})\nabla_i \nabla_j \theta]
  -\frac{1}{2}\int \! d^{d}r \,d^{d}r' a_i({\bf r},t) D_{ij}({\bf r},{\bf r'})a_j({\bf r'},t)\right\},
\end{equation}
%\begin{eqnarray}\label{MSR-action}
%   S&=&\int_{t_{in}}^{t_{f}} dt \{\int \! d^{d}r\, ip({\bf r},t).
%  [\partial_{t}\theta - d_{ij}({\bf r})\nabla_i \nabla_j \theta]    \nonumber \\
%  &&-\frac{1}{2}\int \! d^{d}r \,d^{d}r' a_i({\bf r},t) D_{ij}({\bf r},{\bf r'})a_j({\bf r'},t)\},
%\end{eqnarray}
where we defined $D_{ij}({\bf r},{\bf r'})=d_{ij}({\bf r})+ d_{ij}({\bf r'})-d_{ij}({\bf r}-{\bf r'})$ and
${\bf a}({\bf r},t)=-ip({\bf r},t){\bm \nabla}\theta({\bf r},t)$. In eq.~(\ref{MSR-action}), $p({\bf r},t)$
is an auxiliary field conjugated to the physical field $\theta({\bf r},t)$, enforcing the constraint that
the equation of motion (\ref{Ito-advection}) be satisfied at any time along the trajectories contributing
to the PDF, and the summation on all possible realizations of the stochastic velocity field at intermediate
times has already been performed. It will be convenient to set $p({\bf r},t) \equiv - i \eta ({\bf r},t)$ and
to consider $\eta$ as real in the following.

Extremization of the action with respect to the configurations of both fields $\theta({\bf r})$ and
$\eta ({\bf r})$ between times $t_{in}$ and $t_{f}$, leads to the following set of coupled dual equations
defining extremal (or instantonic) trajectories~:
\begin{eqnarray}\label{dyn-instanton_1}
    \partial_{t}\theta + V_{i}({\bf r},t).\nabla_{i}\theta
    - d_{ij}({\bf r})\nabla_i \nabla_j \theta &=&0,  \\
    -\partial_{t}\eta - V_{i}({\bf r},t).\nabla_{i}\eta
    - d_{ij}({\bf r})\nabla_i \nabla_j \eta &=&0 \label{dyn-instanton_2},
\end{eqnarray}
where the velocity field ${\bf V}({\bf r},t)$ of components
$V_i({\bf r},t) = \int \! d^{d}r' D_{ij}({\bf r},{\bf r'})a_j({\bf r'},t)$ describes
self-consistently the flow in the moving frame of the instanton. We look for particular solutions
of those equations describing the formation of fronts of the scalar $\theta $ and self-similar
blowing-up of its gradient in finite time. They can be parameterized as
\begin{equation}\label{scaling-ansatz}
    \theta({\bf r},t) = (t_{*}-t)^{\alpha}\,\widetilde{\theta}({\bm \rho}),\;\;
    \eta({\bf r},t) = (t_{*}-t)^{-\alpha-\frac{d}{\gamma}}\,\widetilde{\eta}({\bm \rho}),
\end{equation}
where we defined $t_{*}$ as the (arbitrary) critical time of the blowing-up, the scaling position variable
${\bm \rho}$ as ${\bm \rho}={\bf r}(t_{*}-t)^{-\frac{1}{\gamma}}$, and introduced a scaling exponent
$\alpha$ which is bound to be positive due to the conservation of the scalar. The two scaling functions
$\widetilde{\eta}({\bm \rho})$ and $\widetilde{\theta}({\bm \rho})$ are assumed to be odd with respect
to inversion along the compression axis of the flow.

By plugging the Ansatz (\ref{scaling-ansatz}) in eqs.~(\ref{dyn-instanton_1}) and (\ref{dyn-instanton_2}),
it is straightforward
to check that $\widetilde{\theta}$, $\widetilde{\eta}$ and $\alpha$ obey the same equations
(\ref{equation-orbitales}), (\ref{hamiltonien-effectif}) and (\ref{Vitesse}) as $\psi_{0}$, $\varphi_{0}$
 and $\mu_{N}$, provided the overlap $C=\langle\widetilde{\eta}\,|\,\widetilde{\theta}\rangle$ of the
scaling functions (which is conserved along extremal trajectories) is identified
with $N-1$. For a given $C$, the dynamically selected instanton is the one of smallest scaling exponent
$\alpha$, and its action $S$ takes the expression $-{\cal S}(\alpha)\log\frac{t_{*}-t_{f}}{t_{*}-t_{in}}$
(or $-\gamma {\cal S}(\alpha)\log\frac{r}{L}$ if $L$ and $r$ denote typical length scales of the structure
 at initial and final times), where ${\cal S}(\alpha)$ turns out to be the same function as ${\cal S}(\mu)$
 introduced in eq.~(\ref{exposant-simple}). We conclude that there is indeed a complete merging of zero modes
 and instanton pictures in the large $N$ limit.\\

\begin{figure}
\includegraphics{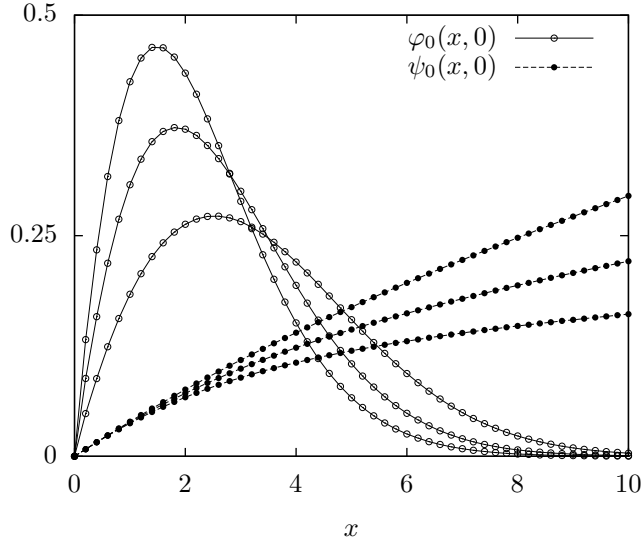}
\caption{\label{champs} Plot of the one-particle orbitals $\varphi_{0}$ and $\psi_{0}$ along
the instanton symmetry axis for $N=4, 8 \;\mbox{and}\; 16$. Curves go from top to bottom and left to right
as $N$ increases. We normalized $\psi_{0}$ by setting its slope at the origin to the constant value 0.04.}
\end{figure}

We solved numerically eqs.~(\ref{equation-orbitales}-\ref{Vitesse}) for $d=2$ and $\gamma =4/3 $.
In order to determine $\varphi_{0}$, $\psi_{0}$ and $\mu_{N}$ for a given value of $N$, we integrate
forward in time the couple of equations $\frac{d|\varphi\rangle}{dt} = l_{N}|\varphi\rangle -
\frac{\langle\psi|l_{N}|\varphi\rangle}{\langle\psi|\varphi\rangle} |\varphi\rangle$ and
$\frac{d\langle\psi|}{dt} = \langle ^{t}l_{N}\psi | - \frac{\langle\psi|l_{N}|\varphi\rangle}
{\langle\psi|\varphi\rangle}\langle\psi|$, until a fixed point is reached. The equations were
discretized within a rectangular box $|x|\leq L_{x}$, $|y|\leq L_{y}$, with $L_{y}=20$ and $L_{x}$
ranging between 20 and 40 according to the value of $N$ (we assume that the particles collapse
onto the $x$-axis, so that the increase of typical lengths with $N$ is more pronounced along
that axis than along transverse directions), and a uniform mesh size $\Delta x = \Delta y = 0.2$.
Some care has to be exercised in the discretizing of fields and differential operators, in order
to get a sensible translation of the condition of incompressible flow on a lattice.
The initial conditions for $\varphi(x,y)$ and $\psi(x,y)$ are arbitrary, except for being odd (resp. even)
functions of $x$ (resp. $y$). As to boundary conditions,
we impose the vanishing of $\varphi$ and the normal derivative of $\psi$ on the perimeter of the box.
We followed the evolution of solutions up to very large values of $N$ (of the order of $10^{6}$).
Figure~\ref{champs} depicts their shape along the positive $x$-axis for $N=4,8$ and 16. It is seen that
bending of the orbital $\psi_{0}$ (the scalar field configuration in the instanton picture) goes
together with an enlargement of the range of the localized dual orbital $\varphi_{0}$. This remains true
at higher values of $N$.

\begin{figure}
\includegraphics{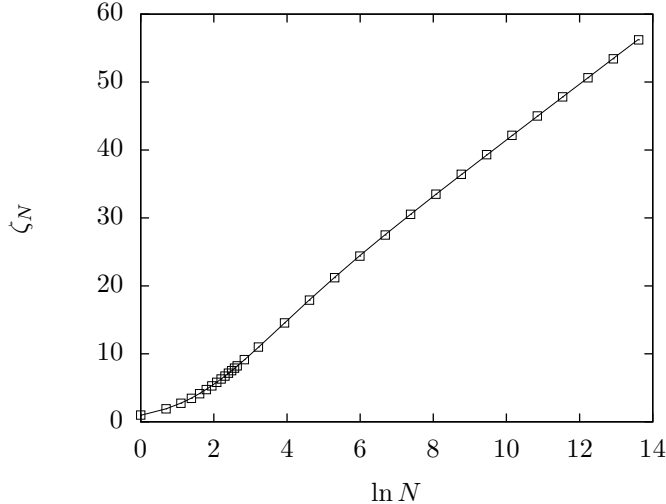}
\caption{\label{results} Numerical estimates of scaling exponents $\zeta_{N}$ (square dots) versus
$\ln N$, obtained for $N$ ranging between 1 and $10^{6}$, with $\gamma = 4/3$ and $d=2$. Data
are almost perfectly reproduced by the analytical expression
$\zeta_{n}=2.87n/(1.0+0.7 \frac{(n+2.00)}{\ln [1.15(n+1.52)]})$ (solid line).}
\end{figure}

The one-particle energy $\mu_{N}$ was found to decrease asymptotically as $1/N$,
so that the action ${\cal S}_{N}$, from eqs.~(\ref{exposant}) and (\ref{exposant-simple}), behaves as
$\log N$ and provides the dominant contribution to the scaling exponent $\zeta_{N}$ at large values of
$N$ (see fig.~\ref{results}). The linear extension of the dumbbell formed by the cloud of particles,
as measured by the distance between the two maxima of $|\varphi_{0}(x,y)|$ along the $x$-axis, turns out
to scale also like $\log N$, while its transverse width increases much slower (like $(\log N)^{\beta}$,
with $\beta$ of the order of $1/3$ for the particular values of parameters investigated numerically).
We lack an analytical explanation of those findings. They show however that particles get more
concentrated as $N$ increases, suggesting that mean field treatment of their
interactions should be relevant.\\

In conclusion, we have shown how the physically motivated assumption of Bose condensation of
Lagrangian particles provides new predictions for the scaling properties of zero modes at large order
and a clear-cut connection between zero modes and self-similar instantons at the same time.
Not surprisingly, the $\zeta_N$ curve shown in fig.~\ref{results} is not quite satisfactory on the
low $N$ side~: in particular, it crosses the dimensional estimate $\zeta_{N}=\frac{\gamma N}{2}$
for $N\simeq 10$ rather than for $N=2$ as it should \cite{FGV01}.
It would therefore be useful to see how quadratic fluctuations (Bogoliubov approximation
in the context of Bose liquid theory) correct the results obtained so far and possibly make them more
accurate even at moderate values of $N$. Note that the inclusion of those fluctuations will restore
some important symmetries in the ground state wave function, like rotational invariance.
Preliminary steps in this direction suggest that this can be done rather harmlessly. Beyond
these quantitative issues, an interesting outcome of this work is that it sets back on stage
the instanton approach as a sound and potentially promising way for capturing, one day, the full
complexity of Navier-Stokes turbulence.

\begin{acknowledgments}
The idea of this work goes back to my participation in the program on ``Developed Turbulence" organized
at ESI in Vienna in 2002 by K. Gaw\c{e}dzki, A. Kupiainen and M. Vergassola. Useful discussions
with B. Castaing and M. Vergassola at early stages of investigations are gratefully acknowledged.
\end{acknowledgments}

%\bibliography{instanton_bib}

%merlin.mbs 2010-03-15 4.21a (PWD, AO, DPC)

\end{document}